# Blockchain Technology: Introduction, Integration and Security Issues with IoT

**Sunil Kumar Singh**

Mahatma Gandhi Central University, Bihar, India

Email Id. sksingh@mgcub.ac.in; sunilsingh.jnu@gmail.com

ORCID iD: 0000-0001-8954-6648

**Sumit Kumar**

Gopal Narayan Singh University, Bihar, India

Email Id. sumit170787@gmail.com

ORCID iD: 0000-0002-4092-385X

**Abstract**

Blockchain was mainly introduced for secure transactions in connection with the mining of cryptocurrency Bitcoin. This article discusses the fundamental concepts of blockchain technology and its components, such as block header, transaction, smart contracts, etc. Blockchain uses the distributed databases, so this article also explains the advantages of distributed Blockchain over a centrally located database. Depending on the application, Blockchain is broadly categorized into two categories; Permissionless and Permissioned. This article elaborates on these two categories as well. Further, it covers the consensus mechanism and its working along with an overview of the Ethereum platform. Blockchain technology has been proved to be one of the remarkable techniques to provide security to IoT devices. An illustration of how Blockchain will be useful for IoT devices has been given. A few applications are also illustrated to explain the working of Blockchain with IoT.

## 1.1 INTRODUCTION

With the emergence of new communication and information technology, security always has been a major concern. In recent, many well-known organizations have faced security breaches. For example, a well popular search engine Yahoo experienced a major attack in the year 2016 resulting in the conciliation of billions of accounts [1]. After doing the security-related research on many companies, it observed that 65% of the data infringement has happened because of a weak or reeved password. Further, it is found that many times sensitive information stealing was done by phishing emails.

Blockchain technology was conceived mainly to address the security issue of cryptocurrency Bitcoin. It has several benefits and is well suited to handle the security issue. In the blockchain system, there is no central database, and it is a kind of system that does not trust the people. This system assumes that anyone can attack on the system, whether part of the system or outsider, can attack the system; therefore, it is a system that is devoid of human consuetude. Moreover, it is enabled with cryptographic features which can be like hashing and digital signature. Blockchain is immutable[1] also, therefore, anyone can store the data. Finally, as many users involved in the blockchain system, changing or adding new blocks in the system needs to be validated by the majority of the users.

Bitcoin is one of the first digital currency [2], created in 2009, underlying blockchain technology. As bitcoin is known as the first cryptocurrency, it was marked as a spire performing currency in the year 2015 and considered a spanking commodity in 2016. Nowadays, besides bitcoin, blockchain is applied in many other areas like Medicine, Economics, Internet of Things, Software Engineering, and many more.

Blockchain technology is getting popular for offering better and foolproof security by removing intermediaries. It also results in reducing the cost of transactions. It is a shared data structure that is amenable for collecting all the transactional history. In blockchain technology, blocks are connected in the form of chains. The beginning block of the blockchain is recognized as the Genesis block [3]. All other blocks are simple blocks. The chain in the blockchain is the link or the pointers connecting the blocks. Blocks, in turn, keeps the transactions that take place in the system.

Many organizations have defined Blockchain technology in different ways. The Coinbase, the bulkiest cryptocurrency exchange across the globe, has established the blockchain as "A distributed, public ledger that contains the history of every bitcoin transaction" [3]. Oxford dictionary bestows a familiar definition stated as "A digital ledger in which transactions made in bitcoin or another cryptocurrency are recorded chronologically and publicly"[4]. Another description is given by Sultan et al., which narrates a very general definition of blockchain technology as "A decentralized database containing sequential, cryptographically linked blocks of digitally signed asset transactions, governed by a consensus model" [4].

Fundamentals of Blockchain technology are supposed to lie in between the 1980s and 1990s of the 20th century [1] though it gained popularity very recently. It is widely recognized in 2008 after the inquisition of cryptocurrency Bitcoin. Blockchain became widely prevalent after the legendary work of Nakamoto [5], though it is a fictitious name and still has not been explored who the actual person is. Nakamoto proposed a technique to replace the centralized architecture with a pear to pear network-based architecture. Initially, blockchain technology was named as two words "block" and "chain"; however, at the end of the year 2016, these two words have been combined to make its blockchain.

Blockchain uses the concept of a ledger which may be seen as a database to maintain the records or a list of transactions. This is similar to the ledger of a hotel. For example, when you check-in in a hotel, the receptionist ask your identity and enters the record in a hardbound register (called ledger). This entry is maintained date and time-wise. One cannot add or remove the entries in between and can only append in the ledger. Thus, the entries cannot be made in between the two entries as well as can't be deleted in between. One can consider the entry as a transaction and pages of the ledger as a block. So it becomes a chain of blocks in the ledger. In case of any eventuality, this hotel ledger is to be consulted for security purposes. Though, this type of ledger is a centralized database of the hotel.

Intermediation is one of the prominent solutions for screening the ownership of assets or transaction processing. Intermediaries' role is to check and validate the participating parties along with the chain of intermediaries. This validation process, apart from time taking, incurs a significant amount of cost. In case the validation fails, it has credit risk too. The blockchain technology promises a way to overcome, representing "A shift from trusting people to trusting math"[6]. i.e., free from human intervention or minimum human involvement.

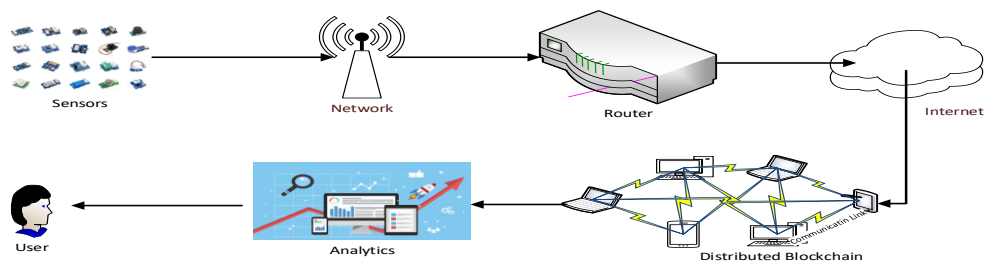

Figure 1.1 Data flow in the IoT-blockchain

Internet of Things is an upcoming technology that indicates the billions of tangible devices across the globe agglutinated to the Internet which collects and shares the information. IoT is the term coined by Kevin Ashton of MIT in 1999 during his work at Procter & Gamble (company)[7]. It promises the world to make it perceptive and proactive by enabling the things to talk with each other [8, 9]. In the Internet of Things (IoT), the collected data from the sensors are maintained in central servers, which may lead to many intricacies when the devices try to communicate with each other through Internet [10]. Centralized locations may also suffer from security issues resulting in their misuses. Blockchain technology can provide a solution in the form of a decentralized model. A distributed model can execute billions of operations between different IoT devices. An IoT with Blockchain has been depicted in figure 1.1, wherein distributed Blockchain replaces the concept of the central server and big data processing at a centralized location. This minimizes the building and maintenance costs associated with the centralized location server. It also reduces the single point failure in the absence of a third party.

This article deliberates on the blockchain and its relevance concerning IoT.

## 1.2 COMPONENTS OF BLOCKCHAIN TECHNOLOGY

Blockchain is a network of blocks (nodes) that are connected with one another following some topology rather than being connected with a central server. It has the potential to store the transactions in the ledger effectively and confirming transparency, security, and auditability. Few crucial components of blockchain technology are as follows.

### 1.2.1 Block

Block in the blockchain technology is the decentralized nodes/miners equipped with the databases, and it contains the digital piece of information. Blocks are linked together containing the hash value of the previous block into the current block. In general, block structure can be visualized into two parts: block header and a list of transactions.

Block header equipped with the following information:

- *Version number* indicates the version number of the block and uses 4 bytes for its representation.
- *Previous block hash* is a pointer between the previous and current block and uses 32 bytes.
- *Timestamp* uses 4 bytes and stores the time of the creation of the block.
- *Merkle tree* is represented by 32 bytes and is a hash of every transaction that takes place in a block.
- *Difficulty target* is indicated by 4 bytes and basically it is used to measure the intricacy target of the block.
- *Nonce* also uses 4 bytes and computes the different hashes.

Figure 1.2 shows a generic diagram of a block with its important components. It also shows the working Merkle root which is generated from the hash values of the transaction. In the given Figure 2, A, B, C, and D are the transactions, and H(A), H(B), H(C), and H(D) are their respective hash values.

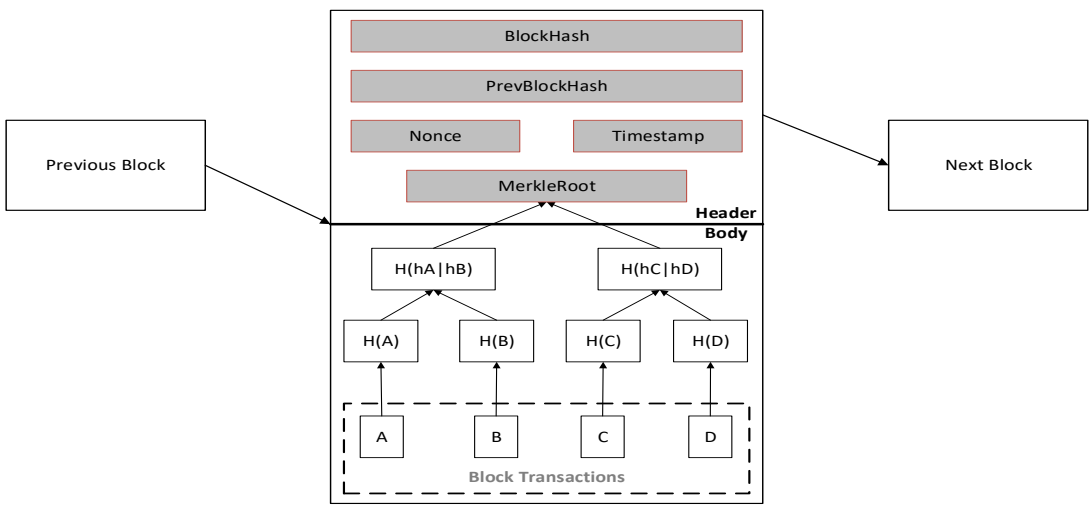

Figure 1.2 Diagram of a Block

### 1.2.2 Genesis Block

In a Blockchain, genesis block is considered as a foundered block because it is the first block in the chain. The Block height of the first block is always zero, and no block precedes the genesis block. Every block which is the part of the blockchain comprises of a Block Header along with Transaction Counter, and Transactions.

### 1.2.3 Nonce

A nonce, an abridgment for "number only used once" is a one-time code in cryptography. Itis a number appended to the hashed (encrypted) block in a blockchain. When it is rehashed, it ensures the difficulty level of antagonism. The Nonce is the number for which blockchain miners solve a complex problem. It is also associated with the timestamp to limit its lifetime; that's why if one performs duplicate transactions, even then a different Nonce is required.

### 1.2.4 User and Miner

A computationally advanced node that tries to solve a complex problem (which requires high computation power) to retrace a new block which is recognized as a miner. The miners are capable of working alone or in a collective routine in order to find the solution to the given mathematical problem. The process of locating a novel block is opened by sharing new transaction information among every user in the blockchain network. It is the responsibility of each user to collect the new transactions into blocks and put their efforts to find the proof-of-work of the block. Proof-of-work is defined as a user is required to solve a computational complex puzzle for publishing a new block and the solution of the puzzle will be its proof. This whole phenomenon is known as proof of work.

### 1.2.5 Chain and Height

In blockchain technology, the chain is a virtual string that connects the miners in the accrescent set of blocks with hashes[11]. The chain keeps growing as and when a new block is appended. Blocks in the chain are generally indicated by their block height in the chain which is nothing but a sequence number starting from zero (0). The height of a block is defined as the number of blocks in the chain between the genesis block and the given block (for which height is to be calculated).

### 1.2.6 Transaction

A blockchain transaction is represented in the form of a smaller unit of the tasks; and is warehoused in public records. After verification by more than 50% of the users of the blockchain network, records get implemented and executed. Its outcomes are stored in the blockchain. Previously stored records can be reviewed at any time but the updations of the records are not permitted. The size of the transaction is a crucial parameter for the miners because the bigger size transaction requires larger storage space in the block. It also requires significantly more power, whereas the smaller size transaction requires less power. The structure of the blockchain [3] is shown in table 1.1.

Table 1.1 Blockchain's Structure

| Field | Size |
| --- | --- |
| Magic Number | 4 bytes |
| Block Size | 4 bytes |
| **Header: Next 80 bytes** | |
| Version | 4 bytes |
| Previous Block Hash | 32 bytes |
| Merkle Root | 32 bytes |
| Timestamp | 4 bytes |
| Difficulty Target | 4 bytes |
| Nonce | 4 bytes |
| **Rest of Blockchain** | |
| Transaction Counter | Variable: 1 to 9 |
| Transaction List | Transaction size-dependent: up to 1 MB |

## 1.3   TYPES OF BLOCKCHAIN

Centred on the uses of blockchain technology for various applications in a different scenario, it is broadly categorized into two categories: Permissioned and Permissionless [1].

### 1.3.1   Permissioned

In this, one required to take some sort of permission from that particular organization or owner of the blockchain to access any or parts of the blockchain. For example, to read a blockchain wouldn't allow us to perform any other operations in the block. One needs to take permission to access or transact the block. Permissioned blockchain is categorized into two categories, as follows.

#### 1.3.1.1  Private Blockchain

The private blockchain is fully permissioned, and if a node is willing to join, it has to be a member of that single organization. This new node needs to send an original transaction and required to take part in the consensus mechanism. The private blockchain is useful and is generally favoured for individual enterprise solutions to record the track of data transfer between different departments [12]. Examples of private Blockchain are Ripple and Hyperledger.

#### 1.3.1.2  Federated Blockchain

Federated blockchain, also known as a consortium blockchain, shares a lot of similarity to a private blockchain. It is a 'semi-private' system that has a controlled user group. A federated blockchain is taken as an auditable and credibly synchronized dispersed database that preserves the track of data exchange information between consortium members taking part in the system. Like a private blockchain, it does not annex the processing fee and incurs a low computational cost to publish new blocks. Federated blockchain ensures the auditability and contributes comparatively lower latency in transaction processing. Examples of federated blockchains are EWF, R3, Quorum, and Hyperledger, etc.

If we compare with the public blockchain (mentioned in sec. 1.3.2.1), private is more comfortable because of less number of users. It requires less processing power and time for verifying a new block. It also provides better security because the nodes, which are within the organization, can read the transactions.

## 1.3.2 Permissionless

A Permissionless blockchain is simple, with no restriction for entry to use it. As the name indicates, anyone and anything can be the part of it without taking permission.

### 1.3.2.1 Public Blockchain

A public blockchain is a Permissionless blockchain in which the validation of transactions depends on consensus. Mostly it is distributed, in which all the members take part in publishing the new blocks and retrieving blockchain contents. Ina public blockchain, every block is allowed to keep a copy of the blockchain, which is used in endorsing the new blocks [12]. A few popular applications of public blockchain execution are cryptocurrency networks which are like Bitcoin, Ethereum, and many others. It has an open-source code maintained by a community and is open for everyone to take part in [1, 13].

A public blockchain is difficult to hack because for adding a new block, it involves either high computation-based puzzle-solving or staking one's cryptocurrency. In this, every transaction is attached to some processing fee.

A comparison of various available technologies [11] is shown in table 1.2.

Table 1.2 Comparison of blockchain technologies

|  | **Public Blockchain** | **Private Blockchain** | **Consortium/Federated Blockchain** |
|---|---|---|---|
| **Participation in Consensus** | every node | Solo organization | Some specified nodes in multiple organizations |
| **Access** | Read/write access allowed to all | High access restriction | Comparatively lower access restriction |
| **Identity** | Pseudo-anonymous | Accepted participants | Accepted participants |
| **Immutability** | Fully immutable | Partially immutable | Partially immutable |
| **Transaction Processing Speed** | Low | High | High |
| **Permission Required** | No | Yes | Yes |

## 1.4 SMART CONTRACT

The Smart Contract is the term, introduced by Szabo in 1997 [11, 14], which combines computer protocols with users to run the terms of the Contract. A smart contract is a self-enforcing agreement (an agreement enforced by the party itself) embedded in computer code managed by the blockchain. It is governed by the computer protocols under which the performance of a reliable transaction occurs without the participation of any third parties. The transactions performed under the Smart Contract can be tracked and is irreversible. A smart contract basically consists of the following components: lines of code, storage file, and account balance. It can be created by a node to initiate a transaction to the blockchain. The lines of code i.e. program code are immutable and cannot be moderated once it is created.

Figure 1.3 shows the contract's storage file associated with the miner and stored in the public blockchain. The network of miners is responsible for executing the program logic and acquiring the consensus on the execution's output. Only that particular node (miner) is enabled to hold, access, and modify the data in the blockchain. The contract's code follows a reactive approach, i.e. it is executed whenever it receives a message from the user or any other nodes in the chain. While during the execution of the code, the Contract may access the storage file for performing the read/write operations.

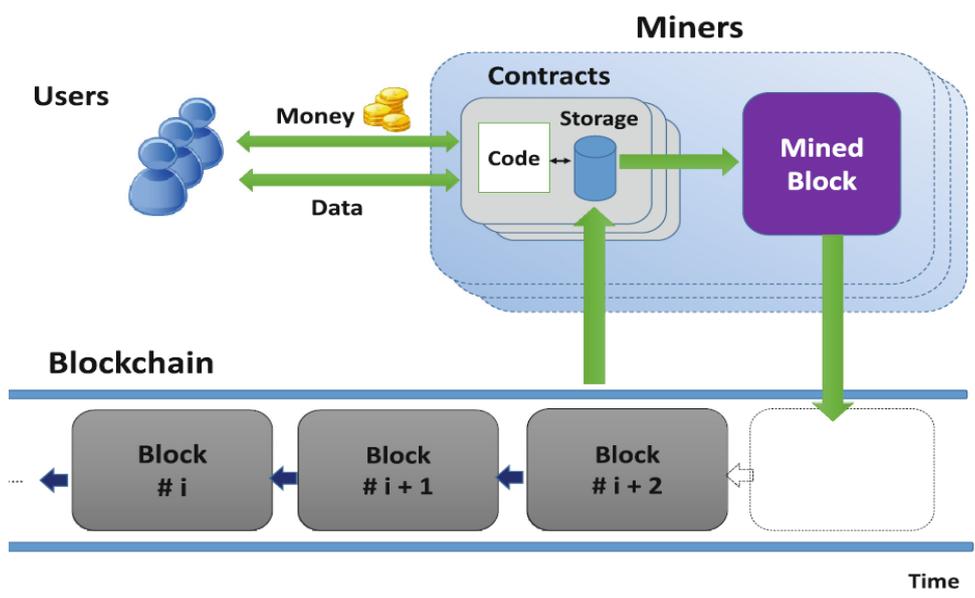

Figure 1.3 Structure of the distributed cryptocurrency system with smart contracts [14]

## 1.5 CONSENSUS MECHANISM

Consensus mechanisms[1] are the protocols that ensure the synchronization of all the nodes with each other in the blockchain. It validates the transaction if it is legitimate before adding it to the blockchain. This mechanism plays an essential character in the smooth and correct functioning of blockchain technology. It also ensures that all the nodes use the same blockchain and all the nodes must continuously check all performed transactions.

Many consensus mechanisms are available today. However, a few known prevalent blockchain consensus mechanisms are Proof of Work (PoW), Proof of Stake (PoS), Delegated PoS, Ripple, and, Tendermint [15]. The key difference among numerous consensus mechanisms can be identified, the way they depute and payoff the authentication of multiple transactions.

Even after the availability of the number of consensus mechanisms, many existing blockchain systems, including Bitcoin and Ethereum uses PoW. PoW is the first and popular consensus mechanism. Its use is widely accepted in many of blockchain-based systems. It is mandatory for the users, participation in the blockchain network, to prove that the work is done for them to qualify and obtain the aptitude to add a new block to the ledger [1]. In the blockchain network, nodes are expected to receive the consensus and agree that the block hash provided by the miner is a valid PoW.

Figure 1.4 illustrates the working of the PoW mechanism in the blockchain. In this mechanism, every miner is first required to define and create a PoW puzzle in the blockchain. The created puzzle will be visible and accessible to every other node taking part in the system. However, the node, which can solve the PoW puzzle, is able to hold, access, and modify the data in the blockchain.

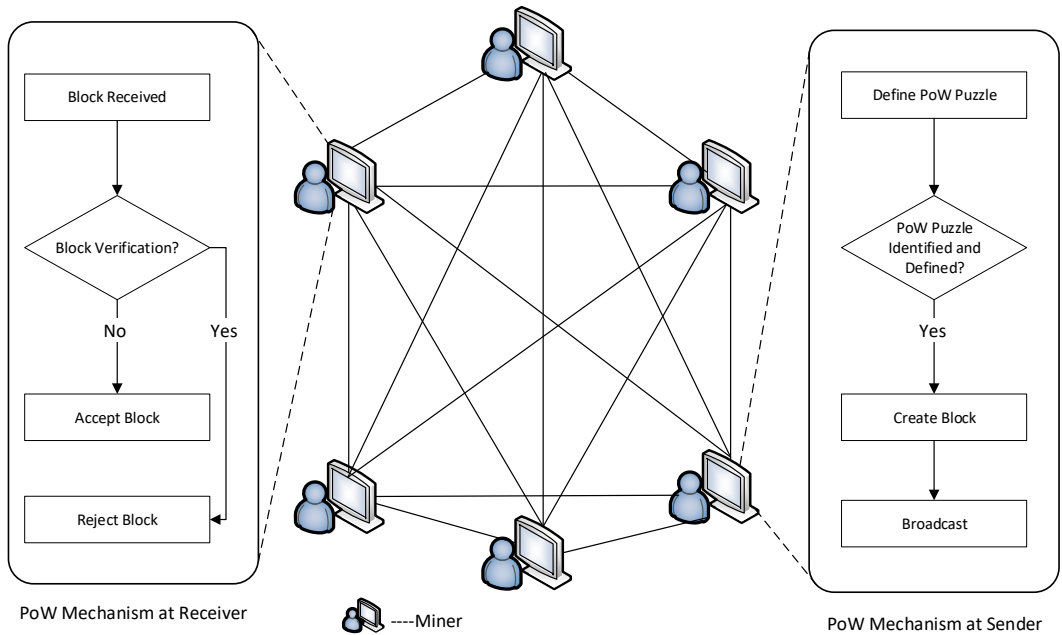

Figure 1.4. Working of Consensus mechanism in blockchain

**1.6   ETHEREUM**

Ethereum is a distributed computing platform, used for public blockchain systems with an operating system featuring smart contract functionalities. It is an open-source platform proposed by VitalikButerin, a programmer and cryptocurrency researcher in late 2013 [16].

Ethereum is also a validated platform used to deliver and execute smart contracts reliably. It supports a modified form of the Nakamoto-consensus mechanism, which works on "Memory Hardness" despite fast computing power machines. Ethereum is a Permissionless network i.e., any node can join the network by creating an account on the Ethereum platform. Moreover, it uses its consensus model, which is identified as EthHash PoW. It is competent to run the scripts using a global network of public nodes. Miner nodes are Ethereum Virtual Machines provided by Ethereum blockchain. These nodes are adequate for providing cryptographic tamper-proof tenacious execution, and its implementation is called contracts. Ethereum reinforces its digital currency known as Ether [17]. Ethereum is one of the well-recognized platforms for executing smart contracts, though, it can execute other decentralized applications and compatible to interact with many other blockchains. It is also categorized as Turing-complete[18], a mathematical concept giving a hint that Ethereum programming language can be used as a platform to simulate other languages.

Ethereum platform may be used to regulate and configure various IoT devices. Security keys are managed using the RSA algorithm, where private keys are stored on the devices, and blockchain controls public keys.

**1.7   BLOCKCHAIN TECHNOLOGY IN IOT**

Blockchain technology can play a vital role for various privacy and security issues of the IoT. In IoT, sensing devices usually send the data at a centralized location for processing purposes. Blockchain technology replaces the central server concept of IoT by introducing the concept of distributed ledger for every transaction with legitimate authentication [10]. It ensures that storing the transaction details with the intermediaries is no longer necessary because transaction records will be available on many computers of the chain. This system rejects the updation and breaching of one computer. However, to make it successful, multi-signature protection is required to authorize a transaction. If a hacker tries to steal the information by penetrating the network, multiple duplicate copies are available on many computers worldwide. For hacking the blockchain network successfully, the consensus of more than 50% of systems in the network is required [19].

### 1.7.1 Blockchain Integration

Integration of Blockchain with IoT opens a new door and wider domain of research and development in the area of IoT applications [16, 20, 21]. Over the last few years, unprecedented growth in the field of IoT has been observed, which enables wide opportunities like access and share of the information. Many times, accessing and sharing information can induce challenges like security, privacy, and trust among the communicating parties. Blockchain can solve various issues of IoT like privacy, security, and reliability. The distributed nature of blockchain technology can eliminate single point failure and makes it reliable.

We all aware that blockchain has already proven its importance in financial transactions with the help of cryptocurrencies, such as Bitcoin and Ethereum. It removed the third-party requirement between P2P payment services [18]. A few IoT enablers have chosen the blockchain technology and formed a consortium for standardization and reliable integration of BIoT (Blockchain-IoT). It is a group of 17 companies aimed to enable security, scalability, heterogeneity, privacy, and trust in distributed structure with the help of blockchain technology [16].

IoT devices can communicate with one another either directly, device to device, or through blockchain technology. There are three types of communication models in an integrated blockchain and IoT environment, which are as follows.

#### 1.7.1.1 IoT-IoT Communication

In IoT-IoT communication, IoT devices communicate directly without the involvement of the blockchain. This type of communication is also known as inter-IoT devices communication. It is the fastest communication model that does not associate high computation and time-consuming blockchain algorithms.

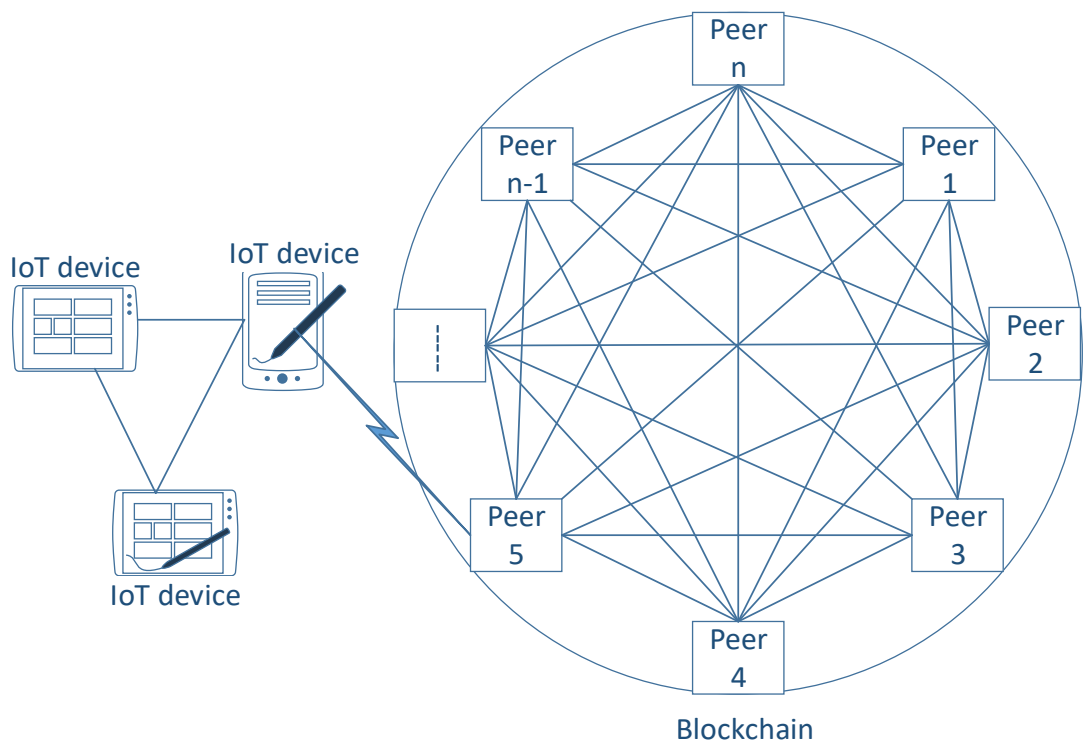

Figure 1.5 IoT-IoT communication model

Figure 1.5 shows that blockchain is not involved in inter IoT communications that's why the system is not able to ensure data integrity, privacy, and security mechanisms. In this model, blockchain stores the communication/transaction history of the IoT devices. This is one of the fast communication models between IoT devices.

### 1.7.1.2 IoT-Blockchain communication

In this model, all transactions among the IoT devices go through the blockchain. This model is enriched with the capability to ensure the data privacy, reliability, and safety of both data and transactions.

Figure 1.6, shows the IoT device communication model through blockchain, which ensures that stored records of each transaction will be immutable, and transaction details are traceable as its features can be verified in the blockchain. Although, blockchain upsurges the autonomy of IoT devices but it may suffer from blockchain overhead which causes latency.

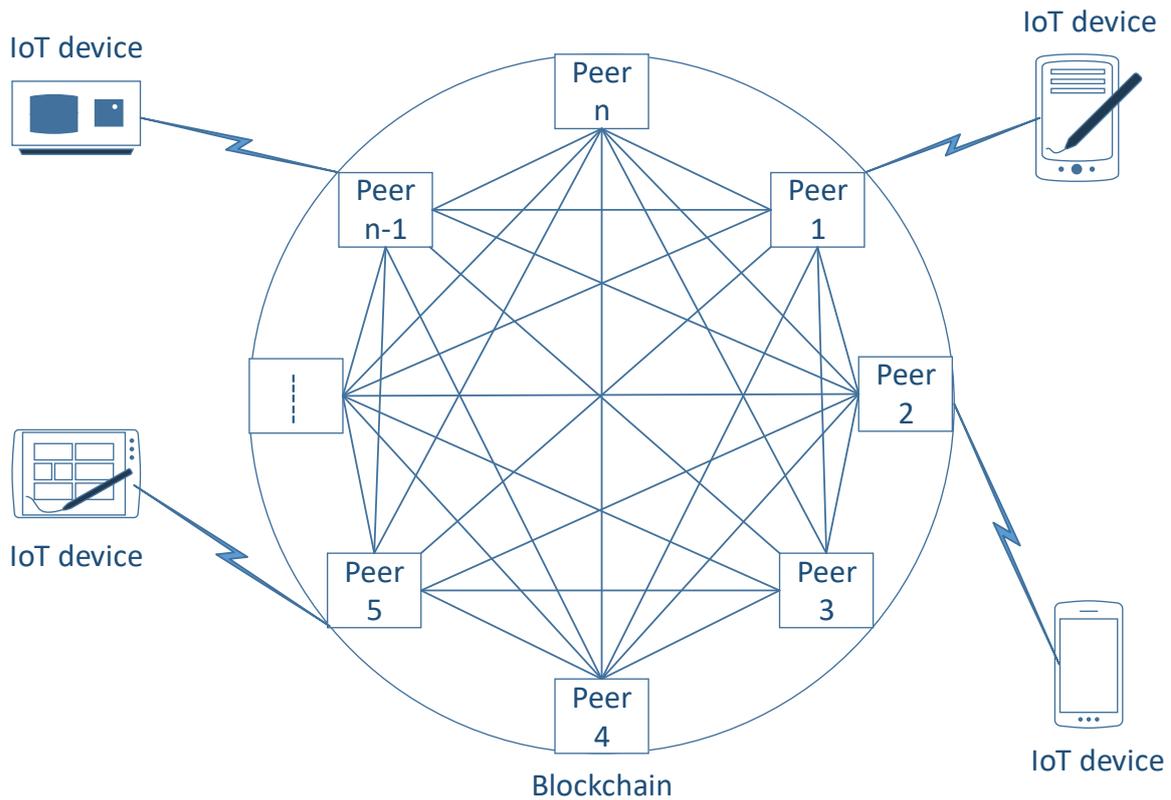

Figure 1.6. IoT-Blockchain communication model

### *1.7.1.3 Hybrid Communication*

The last communication model is a hybrid communication model, in which IoT communication involves the CLOUD/FOG networks. This model shifts partially or most of the computation load, such as encryption, hashing, and compression, from IoT devices to Fog nodes.

Figure 1.7 shows an IoT integrated with blockchain technology, which can transfer high computation load and time-consuming algorithm to Fog node. In this way, fog and cloud computing comes into play and complement the shortcoming of blockchain and IoT [21].

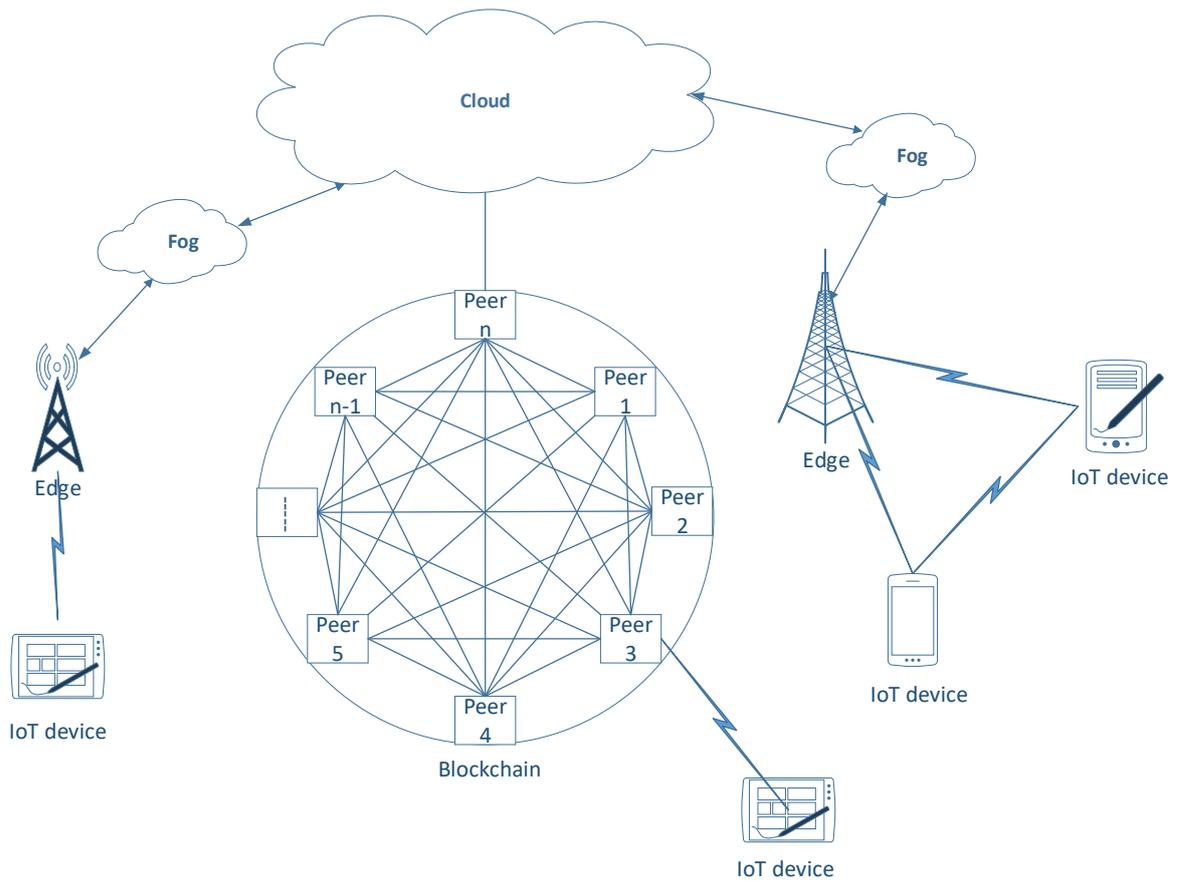

Figure 1.7 Hybrid Communication model

**1.7.2 Security in Blockchain and IoT**

The Internet of Things (IoT) is a structure of Machine-to-Machine (M2M) associations, with no human involvement at all. Hence establishing faith with the participation of machines is a formidable challenge that IoT equipment still has not met broadly. The Blockchain (BC) can take steps as a medium in this process, for improved scalability, protection of data, dependability, and privacy. This process can be done by BC technology to follow all devices which are connected to the IoT environment, and after that, it is used to make possible and/or synchronize all transaction processing. By using the blockchain function, we can fully remove a Single Point of Failure (SPF) in IoT structure. In BC, data is encrypted using various algorithms like cryptographic algorithms and hashing techniques. Therefore, the function of BC provides improved security services in an IoT. The function of BC technology is to repair the digital market. It has a guarantee and retaining both main and preliminary concerns of the function of the BC. The BC keeps the record of a group of sequential and sequence of information transactions since it can be read as a massive networked time-stamping system. The controllers are too concerned in BC's capability to recommend protected, confidential, immediately perceptible monitoring of transactions. Therefore, the BC can facilitate us to avoid the tampering and spoofing of data by the organization and securing the industrial IoT devices [22].

The Blockchain (BC) records every transaction and provides a cross-border overall distributed confidence. Many times, it is possible that Trusted Third Party systems or central location-based services can be vitiated or hacked. In BC, when transactions are confirmed by consensus then the block data are acceptable to all. The BC can be constructed as (1) permissioned network, which is generally a private network, and (2) permission-less, a public network. Permissioned BC offers new privacy and improved access functionality. The BC can resolve these types of challenges effortlessly, strongly, and competently. It has been generally used for providing reliable and certified uniqueness registration, possession track, and monitor of products, supplies, and resources. IoT devices are not exempted, blockchain is able to identity all the connected IoT devices [17].

For security purposes, The BC supports the IoT as mentioned below.

### *1.7.2.1 Data Authentication and Integrity*

The data transmission through IoT devices is linked to the blockchain network and it will be cryptographically proofed and signed via the correct correspondent to hold an exclusive public key and GUID (Global Unique Identifier) which do not require any verification for its uniqueness, and thus it guarantees the verification and truthfulness of transmitting data. Additionally, all transactions complete toward or through an IoT device. Its transaction details are recorded on the blockchain ledger, which enables it to be tracked easily [17].

### *1.7.2.2 Authentication, Authorization, and Privacy*

In BC, smart contracts can offer a decentralized verification policy and sense to be capable of providing a particular and combined verification to an IoT device. The smart contracts are able to provide another effective permission access policy to link IoT devices, employing a smaller amount of complexity while one compares among fixed approval protocols such as Role-Based Access Management (RBAC) [23], OpenID, etc. Nowadays, these protocols are generally used for managing, authorization, and verification of IoT devices [17].

### *1.7.2.3 Safe and Protected Connections*

In general, communication protocols that are used by IoT applications; HTTP, MQTT, XMPP, and many other routing protocols that are not protected in design. These protocols are required to be wrapped with the new security protocols for providing secure communications. The new security protocols can be enriched with blockchain technology; which are DTLS (Datagram Transport Level Security) or TLS [24] with the BC. Key management and identity allocation are completely removed from all IoT devices because it would contain its own single and distinctive GUID and the asymmetric key pair values once mounted and associated to the BC system [17].

Although the BC provides a strong approach for protected IoT, the consensus method depending on the miner's hashing control can be conceded, thus permit the hacker/attacker to host the BC. Also, the private keys among restricted uncertainty can be dried up to compromise the BC accounts. Efficient methods up till now need to be distinct to make sure the privacy of transactions and keep away from race attack, which can affect inside dual spending throughout the transactions [17].

The IoT device has very limited storage and computing power; it can still produce safe and protected keys. Once a key is created, the public key is attached to the Public Key data field in addition to the elected IoT receiver and mined with the BC. While protected data communication via BC is not suggested because access to all nature of a broadcast BC on a server, a BC-based public key swap permitted for IoT to set-up Non-Interactive Key managing Protocols [25]. With a Non-Interactive Protocol, session key series utilizing a mixture of BC data fields as 'salt' may provide an effective solution for updating the IoT session keys for safe and secure data transfer. Still, this research field is required to be explored further for better outcomes [26].

A few other aspects of blockchain technology are; it can resolve the IoT security issues considering its limited storage and low computation power. Because efficient, lightweight, integrated blockchain (ELIB) [27] with IoT devices, protects it from security breaches. ELIB easily copes with the computational complexity and several other issues like low bandwidth, delay, and overhead, etc.

The BC structure is offering a trustful background used for data storage and access. This structure has two characters. One is Data integrity, and another is Role-based data access characters. In Data integrity, the structure avoids data stored within if it is being altered. In Role-based data access, it is a guarantee that the structure recommends special data access permissions toward different users and IoT devices [28, 29].

Compared to the cloud-based centralized system, BC system is a decentralized system that has a benefit in protecting certain specific attacks (e.g. Distributed denial of service (DDOS) attacks). The BC system does concern with the particular point of failure problem, which can occur in the cloud-based centralized system. The centralized system is typically controlled by a manager. If the hacker/attacker pinch the manager's account, they can randomly change the system data. While we were well-known, the data or conversion in the BC system is altered conflict [29].

Privacy and security are most essential in the IoT environment. Within the cloud-based centralized system, user's data are stored randomly, which can simply be hacked by the attackers/hackers. The BC system can offer the independence service by the public-key cryptography method. Furthermore, communication in the IoT environment accepts the AES encryption algorithm, which is extremely flexible to the resource-constrained IoT mechanism. Access control is also an essential mechanism in the IoT system; the smart contract of the BC system be able to offer this type of security service [28].

Researchers have observed that associating Blockchain with IoT is beneficial to handle security and privacy issues, which can probably transform many industries. It is pertinent to mention here that IoT security has always been a pressing concern. To explain this, let us take an example of six IoT devices; the Chamberlain MyQ Garage, the Chamberlain MyQ Internet Gateway, the SmartThings Hub, the Ubi from Unified Computer Intelligence Corporation, the Wink Hub and the Wink Relay; that are tested by a US-based application security company "Veracode" in 2015. The Veracode team found five devices, out of six, had serious security issues. The team was responsible for observing the implementation and various security issues of the communication protocol used in IoT systems. The front-end (services between user and cloud) and back-end (between IoT devices and cloud) were examined, and it is found that except SmartThings Hub, the devices even unsuccessful to have a robust password. Besides, Ubi is deficient in encryption for user connection. These security breaches can cause to a *man-in-the-middle* attack. When the team examined the back-end connection, results were even worst. They also lacked the protection from replay attack.

In the era of technological automation, hacking of IoT devices has severe consequences. The incorporation of blockchain technology in IoT[8] is being well adopted through a broad perspective of measures purported to reinforce security. Narrow Band Internet of Things (NB-IoT) is one of the novel types of IoT which is built on cellular networks. It can directly be deployed on Long-Term Evolution (LTE) architecture. Blockchain technology is applied [30] to ensure reliable data integrity and authentication.

Several mission-critical [31] applications, moving towards automation, are getting popular. Ocado, an online supermarket in Britain, is fully equipped with IoT to stringently improve the warehouse. Installed RFID chips into the Ocado warehouse can sense when the new stock requirements are to be ordered or the status of the remaining number of items in the warehouse. Blockchain technology is used to ensure data integrity, and its decentralized replication technique alienates the requirement to have entire IoT data collected at a central location. This is possible because smart contracts, stored on the blockchain, wouldn't allow any modification to the contracts.

Although blockchain technology can protect from vulnerabilities, it still suffers from some issues. Smart contracts in blockchain are visible to all the users that can cause bugs and vulnerabilities; these are the bugs that can not be fixed in the stipulated time duration. Some other drawback includes its complexity, high computation, and sometimes resource wastage.

## 1.8 A COMPARATIVE STUDY

This section includes a comparative study on the previously developed system with the blockchain-IoT based system. Blockchain technology and IoT can be considered as emerging realities in the current epoch, and these two technologies can transform civilization at a rapid pace [32]. From table 1.3, one can see that wireless sensor networks [33] and the Internet of Things based on systems are not immutable, IoT-Cloud is partially immutable. At the same time, IoT-blockchain is a completely immutable system.

IoT-Cloud allows participant to participant (P2P) sharing while IoT-Blockchain supports P2P as well as a participant to machine (P2M) and M2M sharing also. All other systems support limited sharing only. Table 1.3 lists the blockchain-enabled IoT system with respect to certain properties.

Table 1.3 A comparison of blockchain-IoT based system with traditional SHM systems

|   | **Simple [34]** | **WSN [33, 35]** | **IoT [36, 37]** | **IoT-Cloud [30, 38]** | **IoT-Blockchain [32]** |
|---|---|---|---|---|---|
| 1. **De-centralize** | Completely centralized | Completely centralized | Completely centralized | Mostly decentralized | Entirely decentralized |
| 2. **Reliability** | Highly not reliable | High data tempering | Data tampering is possible easily | Data tampering is possible easily | Tempering is not possible |
| 3. **Storage, Privacy, Security, and Confidentiality** | Low | Low | Intermediate | Intermediate | Considerably Higher |
| 4. **eImmutable Behaviour** | Not immutable | Not immutable | Not immutable | Partially immutable | Fully immutable |
| 5. **Real-time** | Nearly-real time | Real-time | Real-time | Real-time | Nearly-real time |
| 6. **Communication and Transparent information Sharing** | Confined Monitoring | Confined Monitoring | Data processing and monitoring | Data processing, Monitoring and P2P information sharing | P2M and M2M communications, Autonomous decision making using Smart contract-based analysis |
| 7. **Interoperability** | Lower | Lower | Intermediate level | Intermediate level | High |
| 8. **Re-active maintenance** | Lower | Low | Medium | Medium | Effectively High |

Observations from Table 1.3 shows that blockchain-enabled IoT system is the most suitable system to ensure reliability, immutability, interoperability, and security etc. as indicated in the table. Therefore, one can conclude that blockchain technology is the most suited technology for IoT enabled systems. A blockchain-based decentralized system is most suitable for IoT networks; which is validated by a study of [39]. In this, a review is done on centralized, distributed, and decentralized systems using various measures like Accuracy, F-score, Detection Rate etc. [40, 41]. Therefore, we can conclude that decentralized blockchain system is the most suitable system of IoT Networks.

## 1.9 CONCLUSION

This article defines the fundamentals of blockchain technology, along with its components. A comparative study of various blockchain technology is also highlighted. Various application areas are mentioned in this article. A blockchain technology, Ethereum, is described that can be used to implement the public blockchain. It ensures the transparency of the information. The importance of blockchain is also explained with the help of the relevant examples.

IoT is an upcoming technology that is being introduced for a smart environment. With such a prevalent environment, security is a measure of concern. This article also introduces how blockchain can be used for security in IoT. Blockchain, being a distributive technology, plays a good role in IoT security. Comparative study section of the article infers the same.